\documentclass[10pt,aps,prl,twocolumn,superscriptaddress,floatfix,longbibliography]{revtex4-1}

\usepackage{graphicx}
\usepackage{amsfonts}
\usepackage{amssymb}
\usepackage{amsmath}
\usepackage{txfonts}
\usepackage{lipsum}
\usepackage[dvipsnames]{xcolor}
\usepackage{wasysym}
\usepackage[colorlinks=true, allcolors=blue]{hyperref}
\usepackage{bbold}
\usepackage{etoolbox} 
\usepackage[normalem]{ulem}
\usepackage{mathtools} 
\usepackage{multirow} 
\usepackage{hhline}
\usepackage{mathrsfs} 
\usepackage{soul}
\usepackage{array}
\usepackage{changes}

\usepackage[inline]{enumitem}

\usepackage{letltxmacro}
\LetLtxMacro{\oldsqrt}{\sqrt}
\renewcommand{\sqrt}[2][\mkern8mu]{\mkern-6mu\mathop{}\oldsqrt[#1]{#2}}

\begin{document}


\title{Pressure Tuning of Electronic Correlations and Flat Bands in CsCr$_3$Sb$_5$}

\author{Maria Chatzieleftheriou}
\affiliation{Institute for Theoretical Physics, Goethe University Frankfurt, Max-von-Laue-Straße 1, 60438 Frankfurt a.M., Germany}
\affiliation{CPHT, CNRS, {\'E}cole polytechnique, Institut Polytechnique de Paris, 91120 Palaiseau, France}

\author{Jonas B.~Profe}
\affiliation{Institute for Theoretical Physics, Goethe University Frankfurt, Max-von-Laue-Straße 1, 60438 Frankfurt a.M., Germany}

\author{Ying Li}
\affiliation{MOE Key Laboratory for Nonequilibrium Synthesis and Modulation of Condensed Matter, School of Physics, Xi’an Jiaotong University, Xi’an 710049, China}

\author{Roser Valentí}
\affiliation{Institute for Theoretical Physics, Goethe University Frankfurt, Max-von-Laue-Straße 1, 60438 Frankfurt a.M., Germany}

\begin{abstract}
CsCr$_3$Sb$_5$ is a newly identified strongly correlated kagome superconductor, characterized by non–Fermi-liquid behavior at elevated temperatures and intertwined charge- and spin-density-wave order below $T_{DW}\approx 54$K. Under external pressure, this order is suppressed and a superconducting phase emerges. 
This phase diagram, which closely resembles that of high-$T_c$ superconductors, together with a kagome flat band near the Fermi level and possible altermagnetic order, has motivated extensive theoretical and experimental investigations.
To better understand how pressure influences the ordered states, we present a systematic study of the evolution of the electronic properties under applied pressure. Performing DFT+DMFT (density functional theory combined with dynamical mean field theory) calculations, we uncover a complex interplay between the redistribution of spectral weight in the flat bands and the strength of electronic correlations under pressure.
Our results further strengthen the interpretation that pressure effectively weakens electronic correlations through enhanced orbital hybridization.
 This, in turn, strongly suggests that superconductivity emerges as a direct consequence of the suppression of the system’s ordered phase.
\end{abstract}

\maketitle


Flat-band systems have emerged as a major focus of research in recent years~\cite{Cao2018,torma2022, Checkelsky_2024, Crpel2024, aoki2025flatbandscondensedmattersystems}, 
driven by their unique potential to host both strong electronic correlations~\cite{hu2023symmetric,datta2023,PhysRevX.14.031045, lee2025coexistingkagomeheavyfermion} due to the quenched kinetic energy within the flat band, and nontrivial topological effects~\cite{torma2022,Peotta2015}. This combination provides an ideal platform for exploring how topology and interactions intertwine to produce novel quantum phenomena, including
  charge ordering and unconventional superconductivity~\cite{Cao2018, PhysRevB.110.L041121, cualuguaru2022general,kang2020topological, mielke2022time, Zhang2023_NanLett, PhysRevLett.129.166401, PhysRevB.106.115139, Zheng2022_Nat}. Kagome materials provide a natural platform for this physics, as their geometry produces Dirac points, van Hove singularities, flat bands, and nontrivial topology~\cite{mazin2014,PhysRevB.86.121105, Hu2022, Hu2022_NatComm, yin2022topological,ferrari2022charge,Kang2024, Wang_2025} in various combinations across different compounds~\cite{Lou2024, PhysRevMaterials.5.034801, Yin2021, Zhao2021_Nat, Guo2024,lee2025coexistingkagomeheavyfermion}. 

CsCr$_3$Sb$_5$ has recently been identified as a novel kagome material that, owing to its crystal structure and electron filling, features flat bands near the Fermi level and strong orbital-selective electronic correlations~\cite{Hu2023, liu2024superconductivity, wang2025spin}.
At ambient pressure, the system is a correlated bad metal, exhibiting simultaneous charge and spin density wave order below $\sim54 K$~\cite{li2025electron,wang2025spin, peng2024flatbandsdistinctdensity, PhysRevB.110.165104, huang2025controllingaltermagneticspindensity, cheng2025friezechargestripescorrelatedkagome,li2025exoticsurfacestripeorders}.
Under the application of hydrostatic pressure, the two ordered phases separate in temperature~\cite{liu2024superconductivity}. With increasing pressure, these ordered states are eventually suppressed, and a superconducting dome emerges at higher pressures (between $\sim4-10$\,GPa) and low temperatures (below $\sim6.4$\,K)~\cite{wangintriguing,wu2025flat}. 
At present, neither the nature of the spin and charge fluctuations nor the mechanism underlying the unconventional superconductivity is fully understood.
This combination of a rich phase diagram and strong electronic correlations suggests that CsCr$_3$Sb$_5$ is the long-sought strongly coupled counterpart of the weakly coupled AV$_3$Sb$_5$ family of kagome metals~\cite{Kenney_2021, PhysRevMaterials.3.094407, PhysRevMaterials.5.034801}.

To address these experimental questions, theoretical studies based on density functional theory (DFT)~\cite{PhysRev.140.A1133} have shown that the multi-orbital nature of the low-energy bands arises primarily from the Cr 3-$d$ orbitals~\cite{liu2024superconductivity,xu2025altermagnetic, wang2025spin}, which hybridize with the antimony $p_z$ orbital.
  At the DFT level, under ambient pressure the flat bands are found to be located between $\sim100-300$\,meV above the Fermi level, while the experimental findings suggest a much smaller distance from $E_F$~\cite{wang2025spin}.
This discrepancy is believed to originate from the insufficient treatment of correlation effects within DFT. The relevance of strong correlations is suggested by the bad metallic behavior observed in both Angle-Resolved Photoemission Spectroscopy (ARPES), low temperature specific heat data and resistivity measurements highlighting the need for an accurate treatment of electronic correlations~\cite{li2025electron,liu2024superconductivity}.

\begin{figure*}[t!]
\includegraphics[width=1\linewidth]{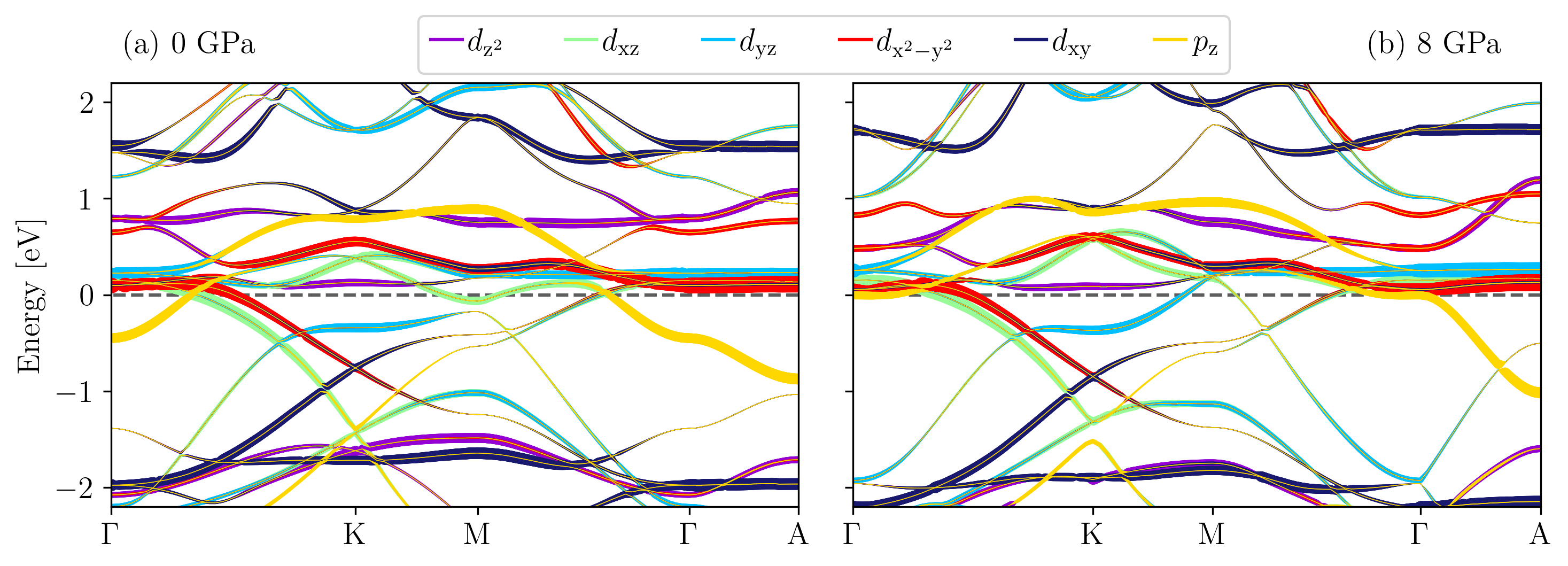}
\caption{The bandstructure of CsCr$_3$Sb$_5$ along the high-symmetry-path for 0 Gpa pressure in panel (a) and for 8 Gpa pressure in panel (b) from DFT. The orbital weights for a the indicated selectoion of orbitals are color-coded by their linewidth. The width of the dots corresponds to the weight of the respective orbital. We observe that pressure induces a reduction in size of the electron pocket near the $\Gamma$ point with largest weight on the Antimony $p_z$ orbital}
\label{fig:bands}
\end{figure*}

A few works have rigorously included the effect of strong electronic interactions in the system, using either the Slave-Spin~\cite{PhysRevB.72.205124, PhysRevB.84.235115, PhysRevB.86.085104, PhysRevB.107.155149} or the Dynamical Mean-Field Theory (DMFT)~\cite{RevModPhys.68.13} approach~\cite{PhysRevResearch.7.L022061,PhysRevB.111.035127,li2025electron,crispino2025tunableelectroniccorrelations135kagome}.
These studies, focusing on the system under ambient pressure, have revealed that correlations indeed shift the flat bands closer to  $E_F$, through a large renormalization of certain orbitals.
These studies further identify the Hund’s exchange as a key factor of orbital-selective mass enhancement~\cite{li2025electron}, placing the material within the framework of Hund’s metal physics, where a decoupling between different orbital characters occurs~\cite{haule2009coherence,georges2013strong}.

The influence of hydrostatic pressure on the electronic structure of CsCr$_3$Sb$_5$
 has not yet been theoretically explored beyond the DFT framework. In particular, the pressure evolution of the flat bands and their potential connection to the suppression of magnetic and charge order, as well as to the emergence of superconductivity, remain open questions. In this work, we address these issues by investigating the multi-orbital nature of band renormalization and spectral-weight redistribution under ambient and finite pressures, incorporating many-body effects within a 
 DFT+DMFT approach. 
\\

{\it Methods} ---
The structure optimization simulations under pressure were performed with the projector-augmented wave method~\cite{Blochl1994, Kresse1999} as implemented in the VASP code~\cite{Kresse1996,Hafner2008}. We used
the generalized gradient approximation (GGA)~\cite{Perdew1996} as exchange-correlation functional and included magnetism. Relaxations were performed assuming a ferromagnetic order of
the Cr atoms. Convergence of the properties of interest was achieved for a 6 $\times$ 6 $\times$ 6 k-mesh and an energy cutoff of 520 eV.
We find that the structural parameters of the experimental ambient-pressure structure ~\cite{liu2024superconductivity}(0 GPa) are closely reproduced by the theoretically relaxed structure at 1.19 GPa (see SM~\cite{SM}), which serves as a reference for theory–experiment comparison. Similarly, the experimental structure at 4 GPa aligns well with the theoretical structure at 5.19 GPa, with the resulting lattice constants reported in SM~\cite{SM}. Considering that the experimental X-ray data were obtained at finite temperature~\cite{liu2024superconductivity}, while our simulations are performed at zero temperature with a specific choice of exchange-correlation functional and magnetic configuration, the level of agreement is remarkably good.

\begin{figure}[t!]
\includegraphics[width=1\linewidth]{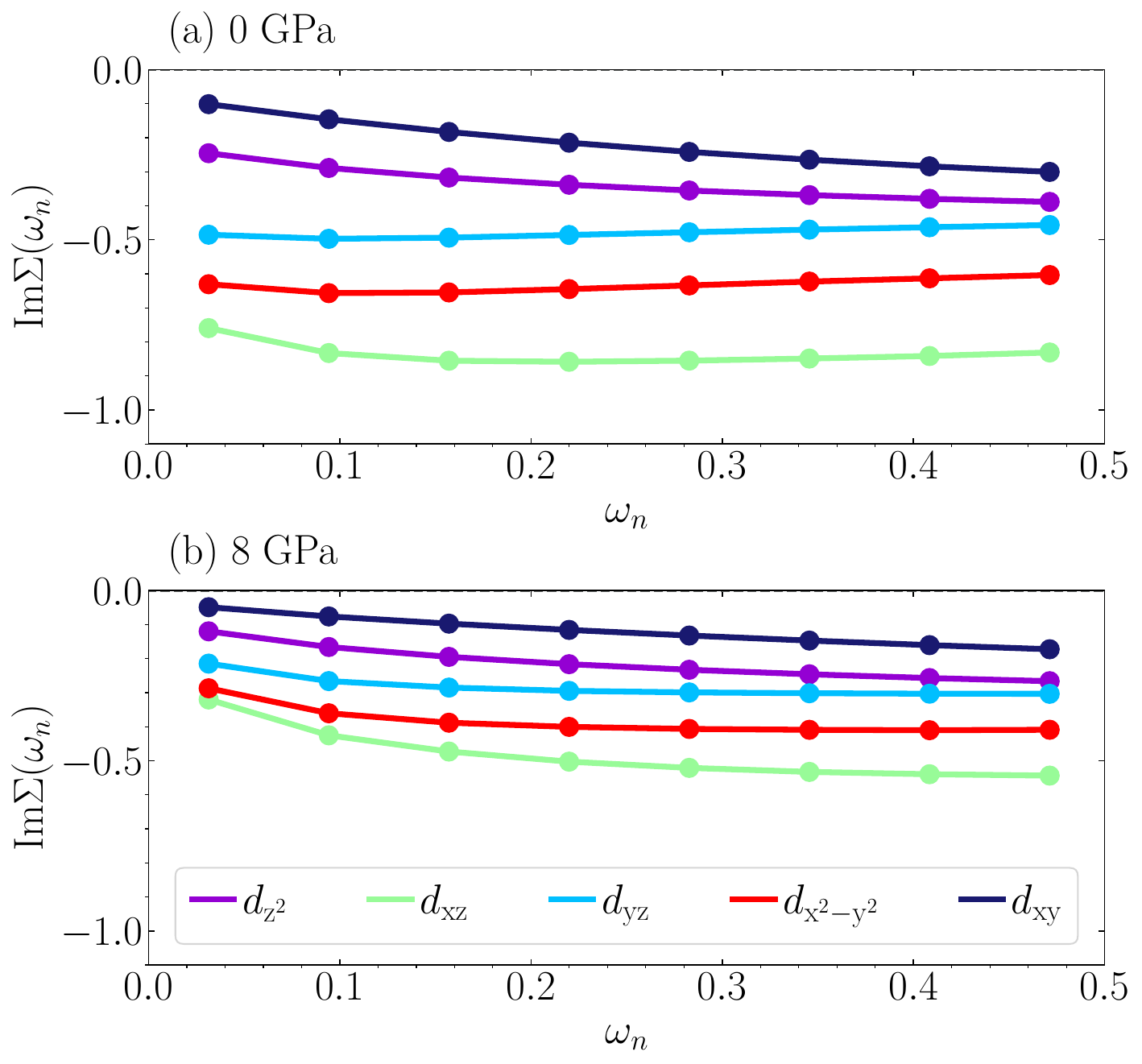}
\caption{Imaginary part of the orbital-resolved self-energy as a function of Matsubara frequencies for pressure $P=0$\,GPa in panel (a) and $P=8$\,GPa in panel (b). The self-energies are overall smaller for a finite pressure, revealing that the system is less correlated. At $P=0$\,GPa we find strong orbital selectivity, with the $d_{xz}$, $d_{yz}$ and $d_{x^2-y^2}$ orbitals being strongly correlated.
\label{fig:ImSigma}}
\end{figure}

In order to estimate the hopping parameters for the structures under pressures, we have performed {\it ab initio} density functional theory calculations with the linearized augmented plane wave (LAPW) method~\cite{Blaha2020}. The Perdew-Burke-Ernzerhof generalized gradient approximation~\cite{Perdew1996} was used, with a mesh of 1000 k points in the first Brillouin zone and $RK_{\rm max}$ was set to 7. The hopping parameters were obtained
through the Wannier function projection formalism implemented in Wien2wannier~\cite{Kunes20101888} and Wannier90~\cite{Pizzi2020}. The calculations were double-checked with the full potential local
orbital (FPLO) code~\cite{Koepernik1999}. 
The valence electrons per unit cell are 34 = 1 × 1(Cs-6s1) + 3 × 6(Cr-3d5,4s1) + 5 × 3(Sb-5p3). Our effective tight-binding model includes 31 = 1 × 1(Cs-s) + 3 × 5(Cr-d) + 5 × 3(Sb-p) orbitals with fixed filling of 34 electrons.
As described in Ref.~\cite{PhysRevResearch.7.L022061}, the dominant orbitals around Fermi energy are Cr $d_{xz}$, $d_{yz}$, $d_{x^2-y^2}$ and Sb $p_z$. As visualized in the SM~\cite{SM}, the dominant hoppings of all these orbitals increase under pressure due to the reduction in interatomic distances.
The DFT bandstructures at ambient conditions and 8 GPa pressure are visualized in Fig.\ref{fig:bands} (a) and (b), respectively. 

For the purpose of accounting for the effect of electronic correlations we employ state-of-the-art Dynamical Mean Field Theory (DMFT)~\cite{RevModPhys.68.13}. 
Within DMFT, we treat the five $d-$orbitals of the three Cr atoms as interacting and all the rest of the orbitals as non-interacting, using a continuous-time hybridization expansion Quantum Monte Carlo impurity solver \textit{w2dynamics}~\cite{Gull_2011, WALLERBERGER2019388}. 
We fix the temperature to ${T=1/100}$\,$e{V^{-1}\simeq116}$\,K, for which the system is at its normal metallic state. 
To determine the interaction parameters within the Cr-3d manifold, we perform constrained RPA calculation~\cite{PhysRevB.70.195104, projcRPA}, see SM~\cite{SM} for details. We obtain the average interaction parameters $U\sim1.5-2.0$\,eV and $J=0.2-0.3U$ in line with previous cRPA calculations~\cite{PhysRevResearch.5.L012008, liu2025electroncorrelationskagomemetals} for other kagome compounds and $d-dp$ type interaction parameters. This choice of interaction parameters is consistent with the choice of correlated $d$-orbitals and $p$ orbitals being projected out on the level of the Green's function~\cite{PhysRevB.70.195104, profe2025exactdownfoldingperturbativeapproximation}. 
We, hence, set the Coulomb repulsion at ${U=2.0}$\,eV and the Hund's exchange coupling at ${J=0.6}$\,eV (in the SM~\cite{SM} we include a discussion on the role of the $J$ value). 
Due to the inclusion of both \textit{d} and \textit{p} orbitals, an unphysical double counting contribution appears, and the relevant term needs to be subtracted from the self-energy. 
To this end, we employ the \textit{fully localized limit (fll)} double counting scheme, which assumes the correlations included in the non-interacting problem to represent the atomic limit of the lattice problem~\cite{WALLERBERGER2019388,PhysRevB.49.14211}.

{\it Results} --- In order to analyze the strength of correlations in the system, we plot in Fig.~\ref{fig:ImSigma} the imaginary part of the self-energy as a function of Matsubara frequencies, for the different Cr $d-$orbitals. The ${P=0}$\,GPa results are shown in panel (a) and the ${P=8}$\,GPa ones in panel (b). At ambient pressure the system is correlated and strongly orbital selective, with the $d_{xz}$, $d_{yz}$ and $d_{x^2-y^2}$ orbitals exhibiting significantly larger mass enhancement and bad metallic behavior compared to the $d_{z^2}$ and $d_{xy}$, in agreement with previous studies~\cite{PhysRevResearch.7.L022061,PhysRevB.111.035127}. The effect of finite pressure (Fig.~\ref{fig:ImSigma}(b)) is to decrease the overall degree of correlations in the material, with all the self-energies becoming smaller and more metallic. In fact, for the less correlated orbitals $d_{z^2}$ and $d_{xy}$, $Im\Sigma\rightarrow0$ in the limit $\omega_n\rightarrow0$, while for the more correlated orbitals $d_{xz}$, $d_{yz}$ and $d_{x^2-y^2}$ in the same limit, $Im\Sigma$ obtains a much smaller absolute value, compared to the situation at ${P=0}$\,GPa.
Moreover, the previously observed strong orbital differentiation gets suppressed correspondingly but does not vanish entirely.

\begin{figure}
    \includegraphics[width=\linewidth]{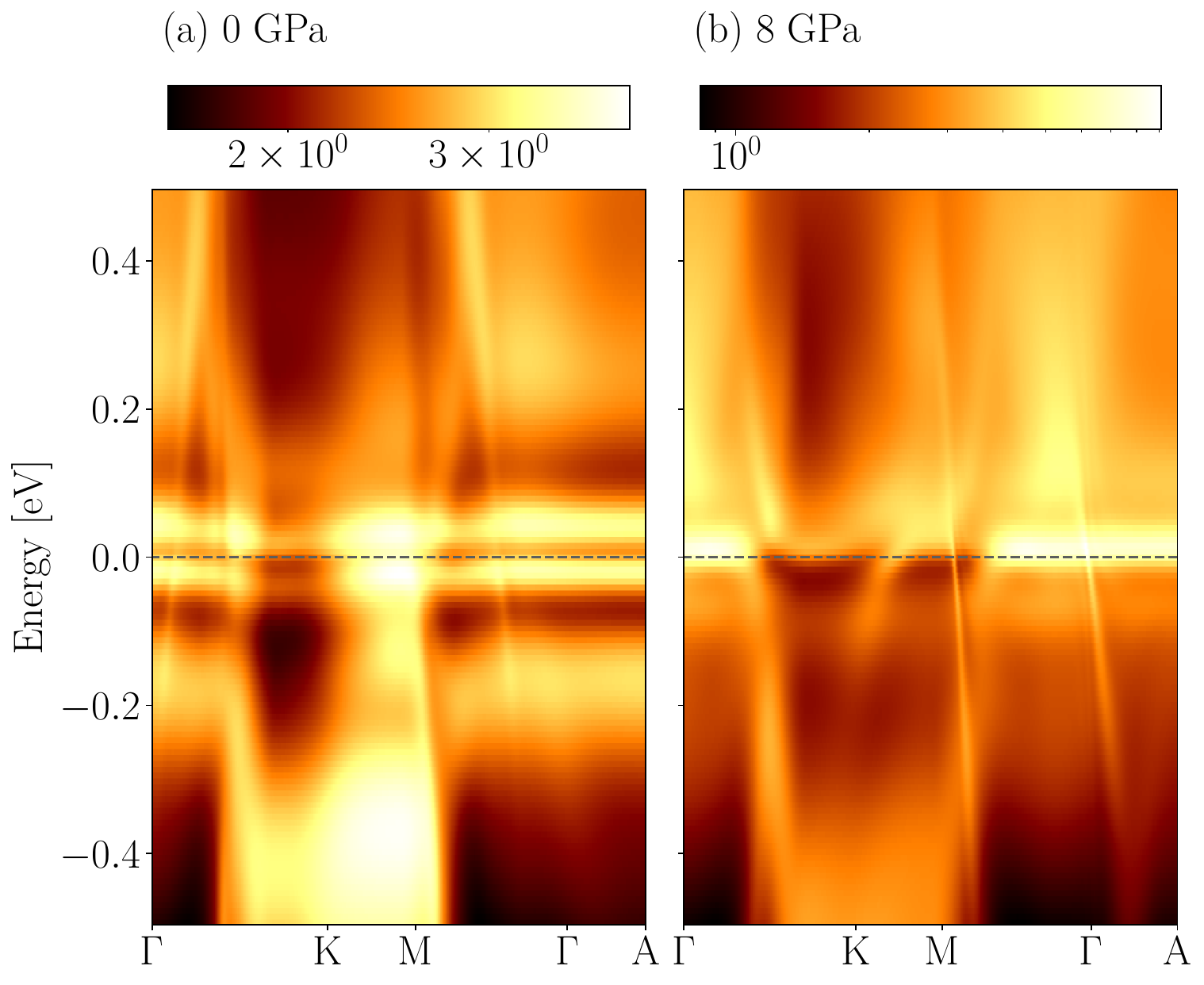}
    \\
    \includegraphics[width=\linewidth]{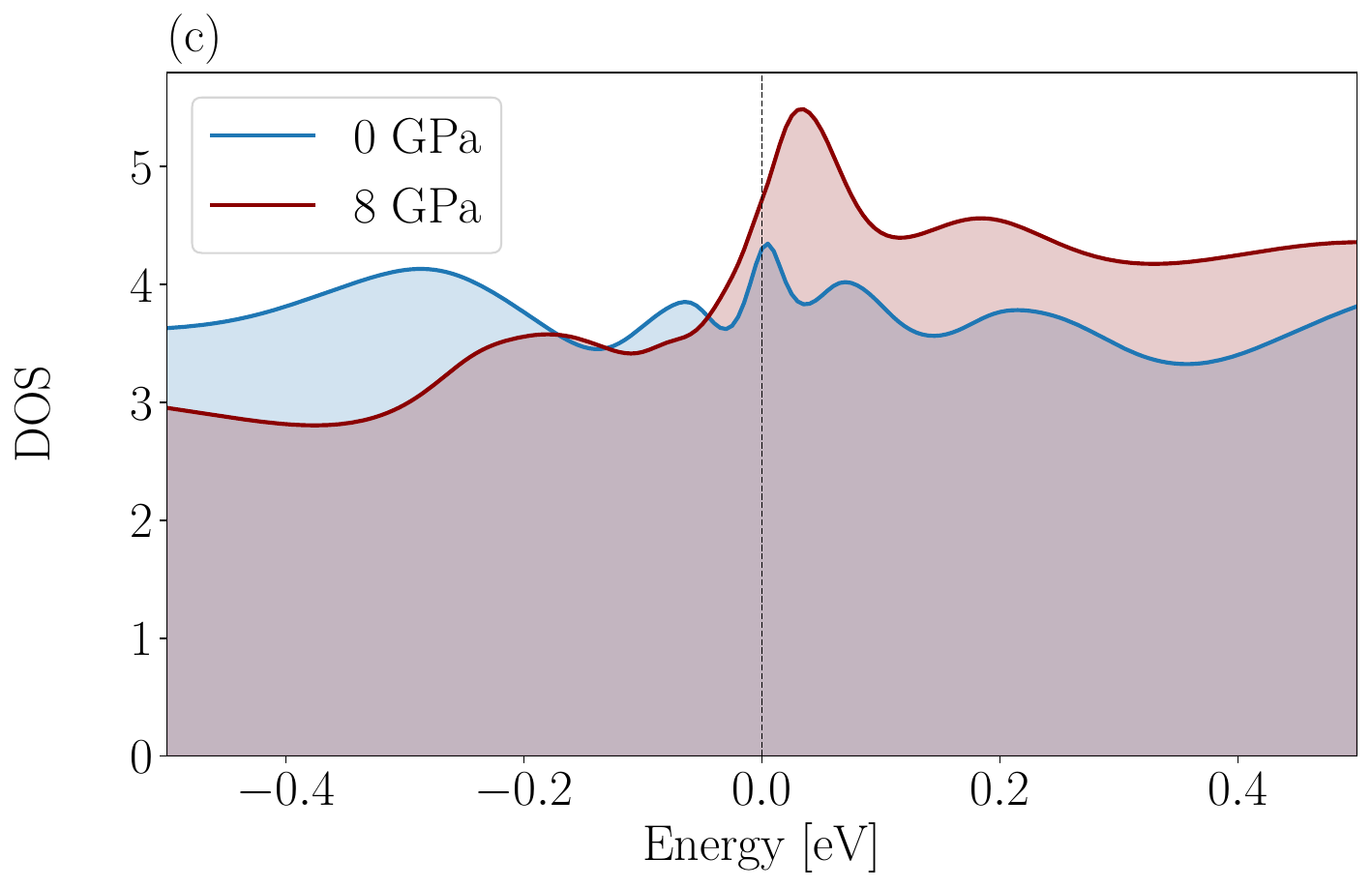} 
    \caption{DFT+DMFT spectral function (in color) as a function of momentum and energy (panel (a) $P=0$\,GPa, panel (b) $P=8$\,GPa), accompanied by the local density of states (DOS) (panel (c)). The spectral function plots reveal that the flat bands are located very close to the Fermi energy ($E_F=0$). At $P=8$\,GPa, weight is redistributed within the flat band, which remains highly weighted only at parts of the momentum path. The DOS at $P=0$\,GPa is more stronly renormalized compared to $P=8$\,GPa, as a direct consequence of the larger degree of correlations, observed from the self-energy.}
    \label{fig:spectral_func}
\end{figure}

We now examine how changes in the correlation strength between the two cases are manifested in the spectral properties of the system.
In Fig.~\ref{fig:spectral_func} the momentum-resolved spectral functions corresponding to ${P=0}$\,GPa (Fig.~\ref{fig:spectral_func} (a))
and ${P=8}$\,GPa (Fig.~\ref{fig:spectral_func} (b)) are plotted along the high-symmetry path $\Gamma$-K-M-$\Gamma$-A as a function of energy, accompanied by the local density of states (DOS) in Fig.~\ref{fig:spectral_func} (c)).  
At ambient pressure (${P=0}$\,GPa) the large correlations lead to a flat band, which is located almost at (but still slightly above) $E_F$, in contrast to its $\sim100-150$\,meV distance from $E_F$ in DFT. A second flat band with reduced weight is found slightly below $E_F$, in agreement to various ARPES measurements~\cite{li2025electron,wang2025spin,peng2024flatbandsdistinctdensity}. In fact, the literature includes contradicting experimental and theoretical evidence, depending on the sensitivity of each method, suggesting a flat band above or below $E_F$. In this work, we resolve the possibility of two flat bands appearing in the spectral function, above and below $E_F$ respectively, as observed in experiments\cite{li2025electron,wang2025spin,peng2024flatbandsdistinctdensity}.
The application of ${P=8}$\,GPa hydrostatic pressure induces a large redistribution of weight, due to the reduction of the correlations.
The system appears to be more coherent, with a high density of states within the positive energy flat band, only at parts of the Brillouin Zone.

\begin{figure*}[t!]
    \centering
    \includegraphics[width=\linewidth]{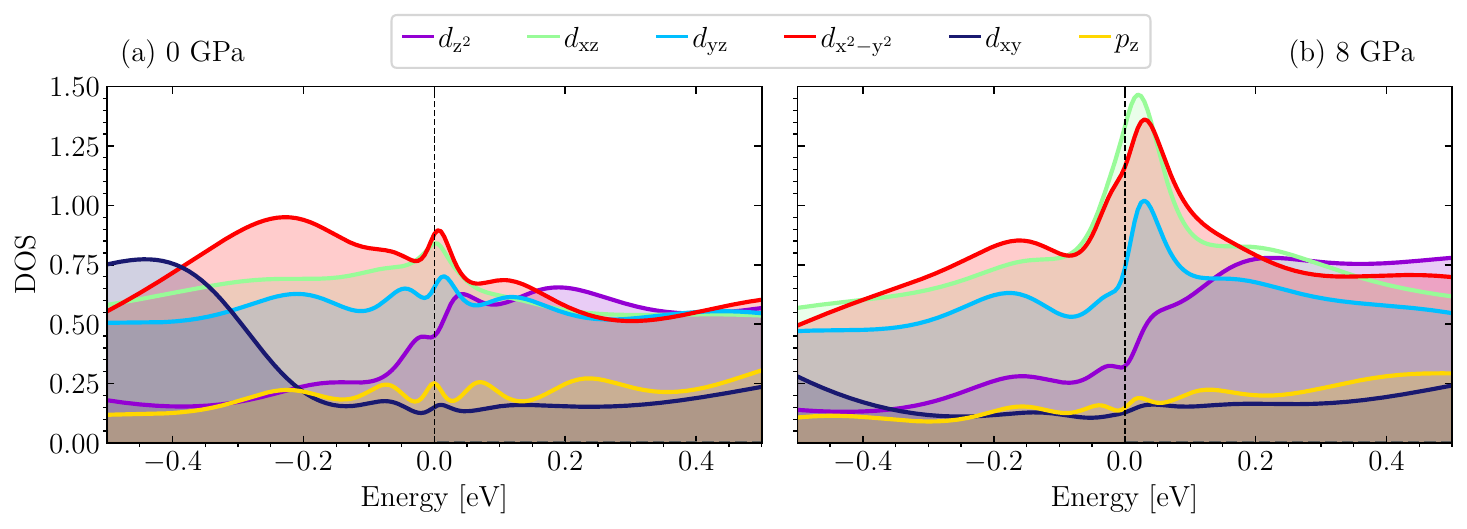} 
    \caption{DFT+DMFT orbitally-resolved density of states (DOS) at $P=0$\,GPa in panel (a) and $P=8$\,GPa in panel (b). The three orbitals that mainly contribute to the flat bands are $d_{xz}$, $d_{yz}$ and $d_{x^2-y^2}$, which are the most strongly correlated orbitals, as illustrated by the self-energy data.}
    \label{fig:DOS_orb}
\end{figure*}

The DFT+DMFT orbitally-resolved DOS, see Fig.~\ref{fig:DOS_orb}, reveals that the main contribution to the flat bands around $E_F$ comes from the $d_{xz}$, $d_{yz}$ and $d_{x^2-y^2}$ orbitals, which are the most strongly correlated ones, as discussed above. Importantly, the peaks corresponding to those orbitals are a few meV away from $E_F$, for both values of pressure ${P=0}$\,GPa, (Fig.~\ref{fig:DOS_orb} (a)) and ${P=8}$\,GPa (Fig.~\ref{fig:DOS_orb} (b)), indicating a pinning behavior near the Fermi-level.
Focusing on the density of states away from $E_F$, one immediately confirms that at zero pressure the system is more correlated, as the high-energy features have more spectral weight, while the low energy quasiparticle peaks get strongly renormalized. 
Further, we find that the Cr-d$_{z^2}$ and the Sb-$p_z$ still contribute a large density-of states in the energy-range in which the flat bands were located in DFT. This is important as scanning tunneling spectroscopy (STS) measurements are mainly sensitive to these orbitals. Therefore, the observed orbital selectivity potentially explains the discrepancy of the strong correlations suggested by the bad metallic behavior in ARPES~\cite{li2025electron,liu2024superconductivity} and a recent STS measurement observing no strong energy shifts of a flat-band compared to DFT~\cite{li2025exoticsurfacestripeorders}.

{\it Discussions} --- In the SM~\cite{SM} the electronic occupancies of the two cases (${P=0}$\,GPa and ${P=8}$\,GPa) are discussed.
The main observation is that both in DFT and in DMFT all orbitals are in proximity to - but not at - half-filling.
The mass enhancement of a correlated metal is often controlled by the orbital occupancies~\cite{Georges2013}. In the regime of strong interactions, if one or more orbitals approach half-filling, even an orbital-selective Mott transition can take place~\cite{Anisimov2002, PhysRevLett.92.216402, PhysRevLett.102.126401}. In that case, the electronic states with those orbital characters become strongly renormalized and localize, behaving effectively as insulating, while the other orbitals remain itinerant and metallic. We observe that the d$_{xz}$ orbital is the one that in DMFT approaches its individual half-filling, by departing from its DFT value. This result is in perfect agreement with the observation that the d$_{xz}$ orbital has the largest self-energy and is the most strongly renormalized one. 

However, we observe that the difference between the occupancies calculated for ${P=0}$\,GPa and ${P=8}$\,GPa is negligible. Thus, this variation cannot explain the large, qualitatively different behavior of the self-energy for $d_{xz}$, $d_{yz}$ and $d_{x^2-y^2}$, among the two values of pressure. 
The culprit behind this qualitatively different behavior can be identified as the spectral weight redistribution within the flat bands and its interplay with the increased kinetic energy due to pressure. 
Therefore, if we want to link the low-temperature ordered phases observed in CsCr$_3$Sb$_5$ to the electronic correlations in the normal state, we must focus on the spectral composition of the flat bands.

At ambient condition, we observe in DFT+DMFT a large, essentially non-dispersive weight at the Fermi-level, indicating a large degree of localization of the electronic degrees of freedom. This agrees with the observation of magnetism in DFT, with a large magnetic moment of $1.7 \mu_{B}$. As mentioned above, under pressure, the DFT+DMFT DOS at the Fermi-level is enhanced, while the flat band becomes more dispersive. This points towards a transition from localized incoherent (zero-dimensional) electronic degrees of freedom to an effectively 2-dimensional (albeit still rather incoherent) electronic system. This change is reflected on the DFT level by a reduction of the magnetic moment to $1.46 \mu_{B}$ at $P = 8$ GPa. 

While we cannot reach the Fermi-liquid regime in our DMFT analysis, these results in combination with prior DFT calculations and experiments reveal a few key aspects for the ordered states~\cite{liu2024superconductivity}. First of all, the clear signature of a structural instability in phonon calculations~\cite{xu2025altermagnetic} indicates that the density wave order is at least partially driven by the phonon sector. When entering a charge-density-wave phase, the Cr kagome lattice is slightly distorted resulting in reduced geometric frustration, thus magnetic order is less penalized and can form. This strong intertwining of magnetism and charge ordering implies that both sectors will be relevant in understanding the superconducting order, as in e.g. the alkali doped fullerides~\cite{Nomura2015}. Our results on the other hand indicate, that pressure reduces correlations and changes the effective dimensionality of the electronic dispersion. Thus, under pressure the material evolves from a regime dominated by localized flat-band physics to one where more dispersive states appear at the Fermi level and become entangled with the flat bands. Overall, this suggests that while flat bands may play a pivotal role in the magnetic state, they are not the primary ingredient for superconductivity. Fully disentangling this interplay will require future studies that approach the ordered phases from a strongly correlated perspective, complemented by alternative routes to tuning correlations—such as the chemical substitution strategy recently explored by Crispino {\it et al.}~\cite{crispino2025tunableelectroniccorrelations135kagome}.


\emph{Conclusion} --- 
We systematically investigated the role of pressure as a tuning parameter in controlling the electronic properties and ordered states of CsCr$_3$Sb$_5$. 
Through DFT+DMFT calculations, we reveal a clear interplay between flat-band spectral weight redistribution and electronic correlations under pressure. Our results show that pressure consistently weakens correlations and orbital selectivity through enhanced hybridization and kinetic contributions, leading to reduced magnetic moments -- as suggested by DFT results -- and the possible suppression of long-range order. These findings strongly support the view that superconductivity emerges from the destabilization of the ordered phase. Together, our work helps reconcile differing experimental results, clarifies the transition from localized to two-dimensional behavior under pressure, and explains key trends in the phase diagram. Moreover, when combined with recent studies~\cite{xu2025altermagnetic, liu2024superconductivity}, our results suggest a delicate and important interplay between electronic and phononic degrees of freedom. 

\begin{acknowledgements}
We would like to thank Fang Xie, Qimiao Si, Matteo Crispino and Giorgio Sangiovanni for useful discussions at the various stages of the work. M.~C., J.~P. and R.~V. acknowledge support by the Deutsche Forschungsgemeinschaft (DFG, German Research Foundation) for funding through projects QUAST-FOR5249 (449872909) (projects P4). The authors gratefully acknowledge the computing time provided to them at the NHR Center NHR@SW at Goethe-University Frankfurt (project number p0026387). This is funded by the Federal Ministry of Education and Research, and the state governments participating on the basis of the resolutions of the GWK for national high performance computing at universities (www.nhr-verein.de/unsere-partner). 
\end{acknowledgements}

\bibliography{Ref}

\end{document}


\title{Supplemental Material\\[0.4cm]
Pressure Tuning of Electronic Correlations and Flat Bands in CsCr$_3$Sb$_5$}

\author{Maria Chatzieleftheriou}
\affiliation{Institute for Theoretical Physics, Goethe University Frankfurt, Max-von-Laue-Straße 1, 60438 Frankfurt a.M., Germany}
\affiliation{CPHT, CNRS, {\'E}cole polytechnique, Institut Polytechnique de Paris, 91120 Palaiseau, France}

\author{Jonas B.~Profe}
\affiliation{Institute for Theoretical Physics, Goethe University Frankfurt, Max-von-Laue-Straße 1, 60438 Frankfurt a.M., Germany}

\author{Ying Li}
\affiliation{MOE Key Laboratory for Nonequilibrium Synthesis and Modulation of Condensed Matter, School of Physics, Xi’an Jiaotong University, Xi’an 710049, China}

\author{Roser Valentí}
\affiliation{Institute for Theoretical Physics, Goethe University Frankfurt, Max-von-Laue-Straße 1, 60438 Frankfurt a.M., Germany}

\begin{abstract}
~\\~\\
\end{abstract}

\maketitle

\begin{figure}[t!]
\includegraphics[angle=0,width=\linewidth]{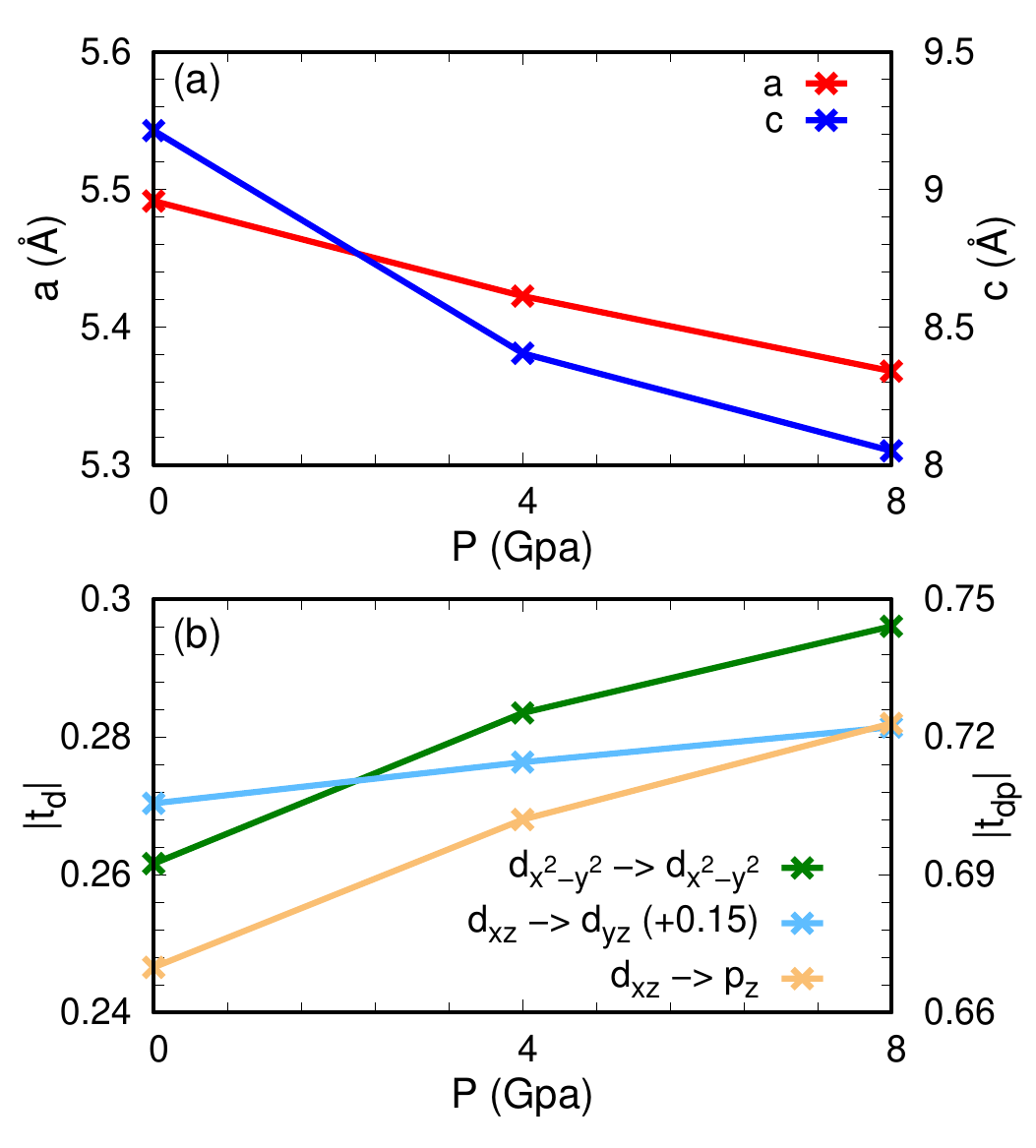}
\caption{Pressure dependence of the (a) lattice constants $a$ (left axis) and $c$ (right axis) as well as (b) dominant hopping parameters within Cr-Cr ($|t_d|$, left axis) and Cr-Sb ($|t_{dp}|$, right axis).}
\label{fig:lathop}
\end{figure}

\section{Details of the DFT calculations}
In Fig.~\ref{fig:lathop}(a) we visualize the dependence of the lattice parameters on pressure. From these calculations we observe that the structural data of the experimental structure~\cite{liu2024superconductivity} at ambient pressure (0 GPa) are rather well reproduced by the theoretically relaxed structure at 1.19 GPa, therefore defining the effective zero of pressure for our theoretical simulations.
The careful relaxation of the structure appears to be crucial for an accurate comparison between theory and experiment. 
As described in Ref.~\cite{Xie2025}, the dominant orbitals around Fermi energy are Cr $d_{xz}$, $d_{yz}$, and $d_{x^2-y^2}$, Sb $p_z$. We therefore plot in Fig.~\ref{fig:lathop}(b) the dominant hopping parameters within these orbitals. All of them increase under pressure due to the reduction in interatomic distances, as expected.

\section{Constrained Random Phase approximation}
\label{app:crpa}
We performed the cRPA calculations with VASP. To this end, we first performed a self-consistent calculation for the relaxed structures on a denser k-mesh of $12\times 12 \times 9$ and an energy cutoff of $600$ eV, with a gaussian broadening of $0.05$ eV. Next, we wannierized the 31 orbital model as described in the main text and performed a projector cRPA with the 15 Cr-d orbitals chosen as the correlated space. To avoid k-point convergence issues from the renormalization of the projectors, we did not consider the long wavelength limit in our calculations~\cite{projcRPA}. We checked that the hierarchy of the values is not altered by the inclusion of the long-wavelength limit and that the broadening is not altering the screening as well. The cRPA interaction parameters for the ambient pressure case are shown in~\ref{tab:placeholder}.

\begin{table}[]
    \centering
    \begin{tabular}{c||c|c|c|c|c}
                     & d$_{xy}$& d$_{xz}$& d$_{yz}$& d$_{z^2}$& d$_{x^2\text{-}y^2}$  \\\hline\hline
d$_{xy}$             & 2.01    &  1.41   & 1.39    & 1.17     & 1.18                   \\
d$_{xz}$             &\J{0.33} &  2.31   & 1.26    & 1.22     & 1.17                   \\
d$_{yz}$             &\J{0.31} &\J{0.51} & 2.24    & 1.16     & 1.21                   \\
d$_{z^2}$            &\J{0.54} &\J{0.47} &\J{0.45} & 2.42     & 1.87                   \\
d$_{x^2\text{-}y^2}$ &\J{0.55} &\J{0.46} &\J{0.47} &\J{0.29} & 2.47                   \\
    \end{tabular}
    \caption{Interaction parameters for the ambient pressure case as extracted from cRPA. We give the Density-density elements of the coulomb tensor in the upper triangular parts and Hund's coupling values $J$ in the lower triangular part in blue.}
    \label{tab:placeholder}
\end{table}

\section{Orbital occupancies}
\label{app:orbocc}

In Table~\ref{tab:occ} the occupancies of the two cases (${P=0}$\,GPa and ${P=8}$\,GPa) are listed extracted both from LDA and DMFT. Here, we observe that in DFT all orbitals are in proximity to - but not at - half-filling, which does not change drastically when treating correlations within DMFT. Therefore, the systems are in the regime in which Coulomb repulsion is not the only dominant interaction term, and Hund's physics becomes important~\cite{citation-0,PhysRevLett.130.066401}.
We observe that the most strongly correlated d$_{xz}$ orbital (it has the largest self-energy as discussed in the main text) is the one that in DMFT approaches its individual half-filling, by departing from its LDA value, both for ${P=0}$\,GPa and ${P=8}$\,GPa. Overall, the difference in the orbital occupancies between the two pressure values are very small.

\begin{table}[t!]
    \centering
    \begin{tabular}{|c|c|c|c|c|}
     \hline
     m & $n_{m}^{LDA}(0$\,GPa) & $n_{m}(0$\,GPa) & $n_{m}^{LDA}(8$\,GPa) & $n_{m}(8$\,GPa) \\
     \hline
    $d_{z2}$ & 0.485 & 0.477 & 0.485 & 0.479 \\
     \hline
    $d_{xz}$ & 0.553 & 0.515 & 0.557 & 0.520 \\
     \hline
    $d_{yz}$ & 0.390 & 0.437 & 0.401 & 0.439 \\
     \hline
    $d_{x^2-y^2}$ & 0.405 & 0.459 & 0.412 & 0.453 \\
     \hline
    $d_{xy}$ & 0.608 & 0.575 & 0.608 & 0.584 \\ 
     \hline
     tot & 4.882 & 4.927 & 4.926 & 4.951 \\
     \hline
    \end{tabular}
    \caption{Orbital occupancies for $P=0$\,GPa and $P=8$\,GPa, obtained by DFT (second and fourth column) and DFT+DMFT (third and fifth column). The value of pressure does not affect significantly the individual orbital occupancies, which remain in proximity to half-filling.}
    \label{tab:occ}
\end{table}

\label{app:hund}
\begin{figure}
    \includegraphics[width=\linewidth]{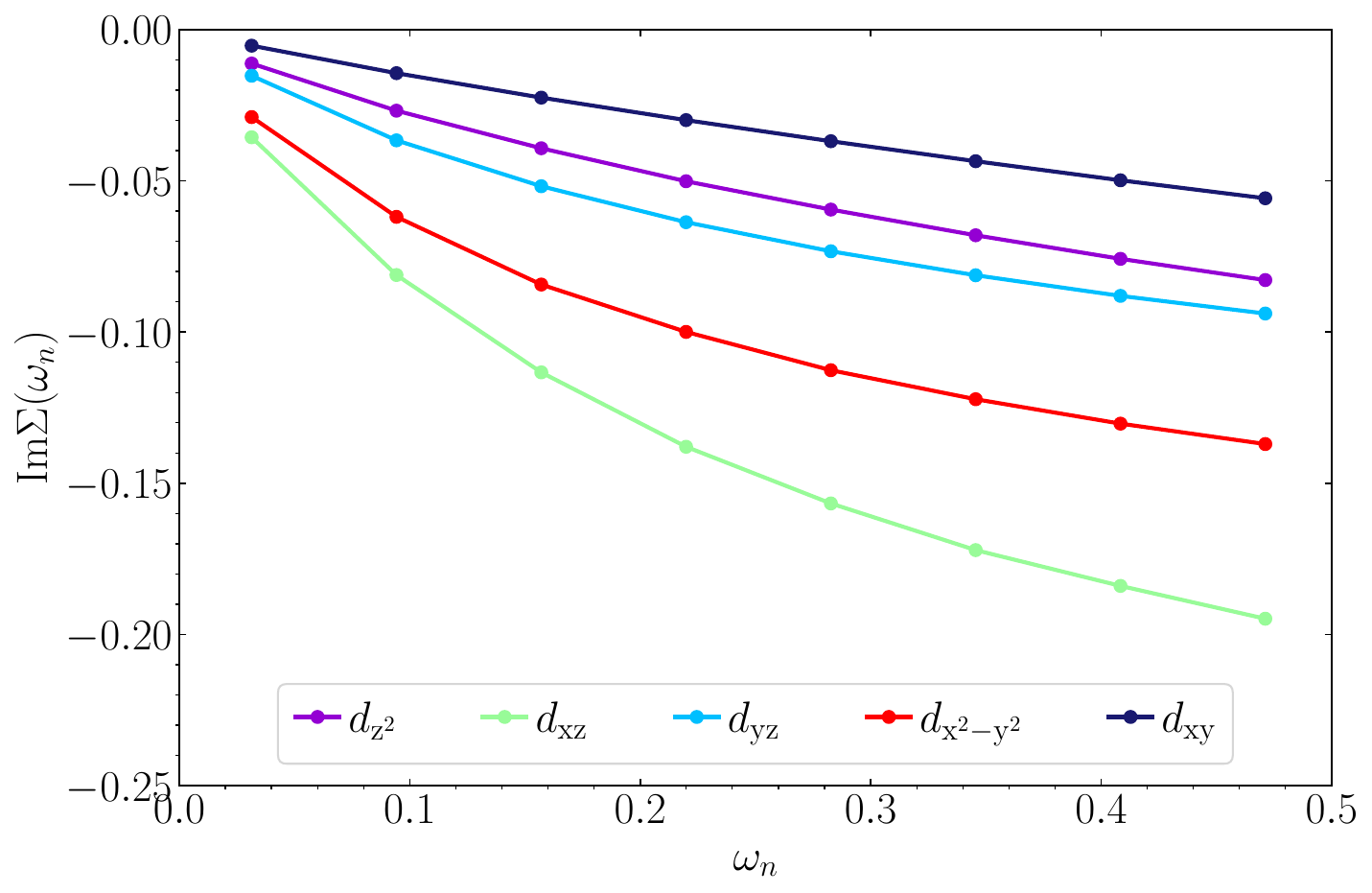}
    \\
    \includegraphics[width=\linewidth]{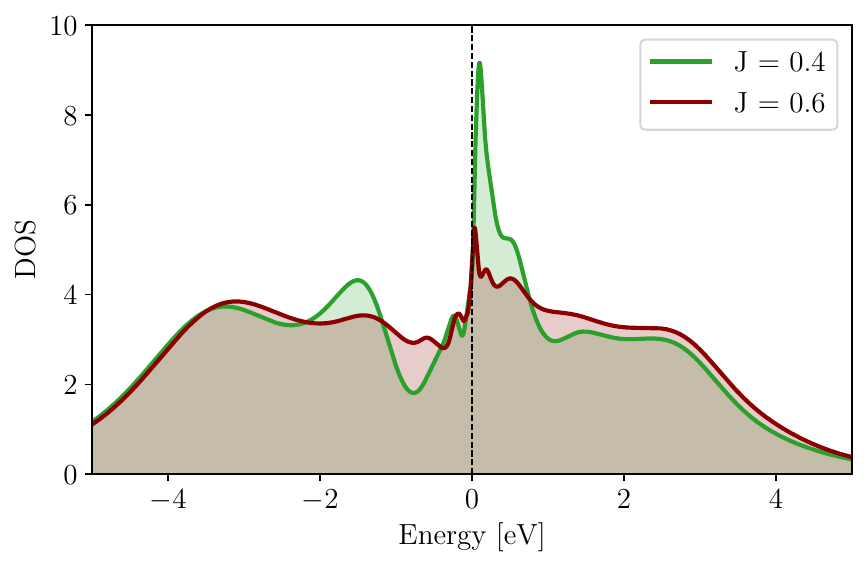} 
    \caption{Results of the system under ${P=8}$\,GPa pressure, with a Hund's coupling $J=0.4$\,eV. \textbf{Upper panel:} Imaginary part of the self-energy per correlated orbital. The correlation strength is suppressed compared to $J=0.6$\,eV, discussed in the main text. \textbf{Lower panel:} Total DOS of the system with $J=0.4$\,eV, compared to the one with $J=0.6$\,eV. The smaller self-energy gives rise to a larger peak around the Fermi energy without a general change in the form of the DOS.}
    \label{fig:J0.4}
\end{figure}

\section{Dependence on Hund's exchange coupling}

As discussed in the main text, by performing cRPA calculations we obtain the average interaction parameters $U\sim1.5-2.0$\,eV and $J=0.2-0.3U$. Throughout this work, we have chosen to focus on the values $U=2.0$\,eV and $J=0.3U=0.6$\,eV. In Fig.~\ref{app:hund} we plot results for a smaller value of Hund's coupling $J=0.2U=0.4$\,eV, for the case of ${P=8}$\,GPa. The imaginary part of the self-energy per correlated $d-$orbital is shown in the upper panel. We observe that the system is significantly less correlated compared to Fig.2 in the main text and, in particular, it is much less orbitally selective. These results are expected, since it is known that Hund's coupling is the term that favors orbital selectivity and enhances correlations. All orbitals exhibit a good Fermi liquid behavior, without alteration in the order of the correlation degree among the orbitals. In the lower panel of Fig.~\ref{app:hund} the total density of states is plotted by comparing the $J=0.6$\,eV and $J=0.4$\,eV cases. We reduction of correlations for $J=0.4$\,eV is reflected in the larger coherent peak around the Fermi level. However, it is interesting to notice that the overall behavior does not change drastically, with the main peak remaining above Fermi, still with distinctive features compared to the ${P=0}$\,GPa case.

\bibliography{Ref}